\documentclass[superscriptaddress,amsmath, nofootinbib]{revtex4}
\usepackage{amsfonts}
\usepackage{graphicx}
\usepackage[latin1]{inputenc}
\usepackage{amsmath}
\usepackage{amssymb}

\begin{document}
%\draft

%\preprint{APS/123-QED}

\title{Comment on "Acceleration of particles to high energy via gravitational repulsion in the Schwarzschild field" by C. H. McGruder III}

\author{Alexei A. Deriglazov }
\email{alexei.deriglazov@ufjf.edu.br} \affiliation{Depto. de Matem\'atica, ICE, Universidade Federal de Juiz de Fora,
MG, Brazil} \affiliation{Department of Physics, Tomsk State University, Lenin Prospekt 36, 634050, Tomsk, Russia}

\author{Walberto Guzm\'an Ram\'irez }
\email{wguzman@fisica.ugto.mx} \affiliation{Depto. de Matem\'atica, ICE, Universidade Federal de Juiz de Fora, MG,
Brazil}

\author{Pablo Rojas }
%\email{pablo.cesar.r.o@gmail.com}
\affiliation{Depto. de Matem\'atica, ICE, Universidade Federal de Juiz de Fora, MG, Brazil}

\date{\today}% It is always \today, today,

\begin{abstract}
By direct computations, we show that "repulsion" in the Schwarzschild field can not accelerate an outgoing particle,
and thus represents  pure coordinate effect. In other words, the repulsion can not be detected neither by local nor by
distant observer.
%We also define three-acceleration that guarantees the impossibility
%for a particle in Schwarzschild field to reach the speed of light.
\end{abstract}

\maketitle %\noindent
%%%%%%{\bf DOI:}
%%%%%%%{\bf PACS numbers:} 11.10.Ef, 03.65.Ca \\
%%%%%%%%{\bf Keywords:}

\section{Introduction}

In a recent paper \cite{Gruder_2017},  McGruder III discussed the radial motion of massive particle emitted at some
point outside the horizon of Schwarzschild field, and then outgoing to infinity. If at the starting point the particle
obeys the condition $\frac{dr}{dt}>\frac{c}{\sqrt 3}(1-\frac{\alpha}{r})$, the geodesic equation implies
$\frac{d^2r}{dt^2}>0$ (see Eq. ({\ref{rep5}) below). The author interpreted this inequality as a gravitational
repulsion, and concluded that the Schwarzschild field can accelerate an outgoing particle up to velocities,
corresponding to the high-energy cosmic rays.

As it was repeatedly pointed out by Spallicci \cite{Spallicci_2017, Spallicci_2010, Spallicci_2016}, in the particular
case\footnote{Initial unrenormalized velocity of such a particle will be above the critical value only if it started
its motion in the region $\alpha<r<3\alpha$.} of a particle that reaches infinity at zero velocity $V_\infty=0$, the
incorrectness of this reasoning follows from the old work of Drumaux \cite{Drum_1936}, and the confusion arises due to
identification of coordinate quantities $\frac{dr}{dt}$ and $\frac{d^2r}{dt^2}$ with velocity and acceleration of the
particle in local environment.  In Sect. 2 we show the incorrectness of McGruder's conclusion for all arrival
velocities: velocity of the particle incoming to infinity always is less than its initial velocity. Since the confusion
around this point\footnote{Detailed history of the subject can be found in \cite{Spallicci_2010}.} has a centenary
history \cite{Droste_1917, Hilbert}, and continues to the present day \cite{Gruder_2017, Loinger_2009, Celerier_2016,
VasIlkov_2016, Gariel_2016, Bini_2017}, we believe that a self-contained presentation of this result could be of
interest.

In Sect. 3 we present Eq. (\ref{rep14}) for acceleration of a geodesic particle in the Schwarzschild field. It also
shows that the field produces an attractive force for any velocity $V<c$. Concerning the expression for acceleration
suggested by Drumaux for the particle with $V_\infty=0$
\begin{eqnarray}\label{rep0}
\frac{d^2R}{dT^2}=-\frac{MG}{r^2}\sqrt{1-\frac{\alpha}{r}},
\end{eqnarray}
our Eq. (\ref{rep14}) implies, that vector of acceleration for that case is given by Eq. (\ref{rep15}), while Eq.
(\ref{rep0}) represents the magnitude of the acceleration, see Eq. (\ref{rep16}). Hence the minus sign in Eq.
(\ref{rep0}) should be corrected. Being always positive, the magnitude (\ref{rep0}) can not be used to say, whether the
Schwarzschild field is attractive or repulsive.

\section{Schwarzschild field can not speed up an outgoing particle}
Consider a massive particle which obeys the geodesic equation $\ddot x^\mu+\Gamma^\mu{}_{\alpha\beta}\dot x^\alpha \dot
x^\beta=0$, written for the trajectory in parametric form $x^\mu(s)$, and it was denoted $\dot
x^\mu\equiv\frac{dx^\mu}{ds}$. The equation implies the conserved quantity $g_{\mu\nu}\dot x^\mu\dot x^\nu=-C_1,
C_1=\mbox{const}>0$. For the radial motion $x^0=ct, x^1=r, \theta=\frac{\pi}{2}, \varphi=0$ in Schwarzschild metric
\begin{eqnarray}\label{rep1}
-d\tau^2=g_{\mu\nu}dx^\mu dx^\nu=-\left(1-\frac{\alpha}{r}\right)(dx^0)^2+
\left(1-\frac{\alpha}{r}\right)^{-1}dr^2+r^2[d\theta^2+\sin^2\theta d\varphi^2], \qquad \alpha=\frac{2MG}{c^2},
\end{eqnarray}
the geodesic equations read
\begin{eqnarray}\label{rep2}
\ddot r=\frac{\alpha}{2r^2}\left[\left(1-\frac{\alpha}{r}\right)^{-1}\dot r^2-\left(1-\frac{\alpha}{r}\right)(\dot x^0)^2\right],
\end{eqnarray}
\begin{eqnarray}\label{rep3}
\frac{d}{ds}\left[\left(1-\frac{\alpha}{r}\right)\dot x^0\right]=0, \quad \mbox{or} \quad (1-\frac{\alpha}{r})\dot x^0=C_2,
\end{eqnarray}
while the conserved quantity acquires the form
\begin{eqnarray}\label{rep4}
(\dot x^0)^2-\left(1-\frac{\alpha}{r}\right)^{-2}\dot r^2=C_1\left(1-\frac{\alpha}{r}\right)^{-1}.
\end{eqnarray}
We are interested in the radial coordinate $r$ as a function of coordinate time $t$: $r(t)$. Using (\ref{rep3}), we
exclude the parameter $s$ from Eq. (\ref{rep2})
\begin{eqnarray}\label{rep5}
\frac{d^2r}{dt^2}=-\frac{\alpha}{2r^2}\left[c^2(1-\frac{\alpha}{r})-3\left(1-\frac{\alpha}{r}\right)^{-1}\left(\frac{dr}{dt}\right)^2\right].
\end{eqnarray}
This is Eq. (3) of McGruder III. The quantity $\frac{d^2r}{dt^2}$ becomes positive above the critical unrenormalized
velocity\footnote{The notions of unrenormalized and semi-renormalized velocities/accelerations have been extensively
discussed in the literature \cite{Thirring, Shapiro_1972, Cavalleri, Spallicci_2010}.}, $\frac{dr}{dt}>\frac{c}{\sqrt
3}(1-\frac{\alpha}{r})$, hence the spellbinding term "gravitational repulsion".

Resolving Eq. (\ref{rep4}) with respect to $\dot x^0$, and using the resulting expression in (\ref{rep3}), we obtain
the  total energy of the particle (see Sect. 88 in \cite{LL})
\begin{eqnarray}\label{rep6}
E\equiv \frac{mc^2C_2}{\sqrt{C_1}}=
mc^3\left(1-\frac{\alpha}{r}\right)^{\frac12}\left[c^2-(1-\frac{\alpha}{r})^{-2}\left(\frac{dr}{dt}\right)^2\right]^{-\frac12}.
\end{eqnarray}
If, instead of this, we resolve (\ref{rep3}) with respect to $\dot x^0$ and substitute into (\ref{rep4}), we obtain an
equivalent expression for the total energy
\begin{eqnarray}\label{rep7}
\left(1-\frac{\alpha}{r}\right)^{-3}\left(\frac{dr}{dt}\right)^2-c^2(1-\frac{\alpha}{r})^{-1}=-c^2\left(\frac{mc^2}{E}\right)^2.
\end{eqnarray}
This is Eq. (5) of McGruder III.

Consider the laboratory $O_r$ fixed at the point $r>\alpha$, and another laboratory $O_\infty$ fixed at the point
$r_\infty$ in asymptotic region where the Schwarzschild metric can be approximated by the Minkowski metric:
$g_{\mu\nu}\rightarrow\eta_{\mu\nu}=(-,+,+,+)$. Let in $O_r$ was created a particle with the magnitude of velocity
$V(r)$, which arrived at $O_\infty$ with the magnitude $V_\infty$. According to Eq. (\ref{rep7}), the conservation law
$E(r)=E(r_\infty)$ implies
\begin{eqnarray}\label{rep8}
\left(1-\frac{\alpha}{r}\right)^{-3}\left(\frac{dr}{dt}\right)^2-c^2(1-\frac{\alpha}{r})^{-1}=
\left.\left(\frac{dr}{dt}\right)^2\right|_{r\infty}-c^2.
\end{eqnarray}
Since in the laboratory $O_\infty$ we deal with the Minkowski metric, distance and time interval coincide with the
coordinate differences $dr$ and $dt$, and we have $\left.\left(\frac{dr}{dt}\right)^2\right|_{r\infty}=V^2_\infty$. At
the point $r$, according to the standard prescription\footnote{We stress that the definition of space and time
intervals between two events in general relativity is not a matter of agreement, but follow from detailed analysis of
the notion of simultaneous events using the null geodesics \cite{LL, Recent_2017}.} (see Sect. 84 in \cite{LL}), the
distance $dl$ and time interval $dT$ are related with the coordinate differences as follows
\begin{eqnarray}\label{rep9}
dl^2=g_{rr}dr^2=\left(1-\frac{\alpha}{r}\right)^{-1}dr^2, \qquad dT=\sqrt{-g_{00}} ~  dt=\left(1-\frac{\alpha}{r}\right)^{\frac12}dt.
\end{eqnarray}
Then the relation between magnitude of velocity and $\frac{dr}{dt}$ is
\begin{eqnarray}\label{rep10}
V(r)\equiv\frac{dl}{dT}=\left(1-\frac{\alpha}{r}\right)^{-1}\left|\frac{dr}{dt}\right|.
\end{eqnarray}
It should be noted that even in one-dimensional curved space, one needs to distinguish the magnitude of velocity $V$
from the velocity vector ${\bf
V}=\frac{1}{\sqrt{-g_{00}}}~\frac{dr}{dt}=\left(1-\frac{\alpha}{r}\right)^{-\frac12}\frac{dr}{dt}$. They are related as
follows: $V=\sqrt{g_{rr}{\bf V}{\bf V}}$. Below we only work with the magnitude of velocity, or velocity for short.

Using these expressions in Eq. (\ref{rep8}), we obtain the following relation between initial and final velocities
\begin{eqnarray}\label{rep11}
\left(1-\frac{\alpha}{r}\right)^{-1}\left(c^2-V^2(r)\right)=c^2-V^2_{\infty} .
\end{eqnarray}
This implies that outgoing particles with initial velocity $V^2<\frac{\alpha}{r}c^2$ can not arrive the laboratory
$O_\infty$. The particles with initial velocity $\frac{\alpha}{r}c^2<V^2<c^2$ will reach the laboratory $O_\infty$ with
the velocity $V^2_\infty<V^2$. Hence the gravitational repulsion implied by Eq. (\ref{rep5}) can not speed up an
outgoing particle, and represents pure coordinate effect. It can not be detected neither by local nor by distant
observer.

This argument could be probably extended to the case of a laboratory at rest in an arbitrary spherically symmetric (in
particular, non static) field\footnote{We are grateful to the referee for pointing out on this possibility.}. According
to Birkhoff  theorem, there is a coordinate system where metric acquires the Schwarzschild form. So the conservation
law (\ref{rep7}) remain valid, and can be used to compare the initial and final velocities of a geodesic particle.

\section{Three-dimensional acceleration in general relativity}

We start with three comments concerning the equation (\ref{rep5}). \par

\noindent 1.  Even in Euclidean space equipped with curvilinear coordinates, the nonvanishing value of
$\frac{d^2r}{dt^2}$ has no the meaning of acceleration. For instance, consider the equation $\frac{d^2x}{dt^2}=0$ for
one-dimensional particle propagating with constant velocity.  In the curvilinear coordinates, say $x=r^2$, this
equation reads $\frac{d^2r}{dt^2}=-\frac{1}{r}\left(\frac{dr}{dt}\right)^2$, the latter can be compared with Eq.
(\ref{rep5}). \par

\noindent 2. After excluding the parameter $s$ from geodesic equations, the formalism remains covariant under
general-coordinate transformations that do not involve the coordinate time \cite{LL, Recent_2017}. For the present
case, they are $r=r(r')$, and one expected that an acceleration should be described by covariant quantity, which is not
the case of $\frac{d^2r}{dt^2}$.
\par

\noindent 3. Since the general relativity is a relativistic theory, we expect that particle in a gravitational field
can not reach the speed of light. As a consequence, the dependence of acceleration on velocity is an inevitable
property of a relativistic theory: three-acceleration in the direction of velocity should vanish in the limit
$V\rightarrow c$. Once again, it is not the case of $\frac{d^2r}{dt^2}$ given by (\ref{rep5}).

In an arbitrary gravitational field, the three-acceleration that obeys the above mentioned properties has been
presented in \cite{Recent_2017}, see Eq. (40) in arXiv:1710.07135. In the recent works \cite{AAD_2018, AAD_2017}, it
was tested in the analysis of a rotating body in general relativity. Being applied to the present case,  Eq. (40)
implies the following expression for one-dimensional vector of acceleration:
\begin{eqnarray}\label{rep12}
{\bf a}=D_T{\bf V}=\frac{dt}{dT}\frac{d{\bf V}}{dt}+\Gamma^r{}_{rr}(g_{rr}){\bf V}{\bf
V}=\left(1-\frac{\alpha}{r}\right)^{-1}\frac{d^2r}{dt^2}-
\frac{\alpha}{r^2}\left(1-\frac{\alpha}{r}\right)^{-2}\left(\frac{dr}{dt}\right)^2.
\end{eqnarray}
By construction, ${\bf a}$ transforms as a vector under residual general-coordinate transformations
$r=r(r')$. Magnitude of this vector, $a=\sqrt{g_{rr}{\bf a}{\bf a}}$, coincides with $\frac{dV}{dT}=\frac{d^2l}{dT^2}$.
Representing $\frac{d^2r}{dt^2}$ in Eq. (\ref{rep5}) through the acceleration, we obtain second law of Newton for the
particle propagating along a geodesic line
\begin{eqnarray}\label{rep14}
{\bf a}=-\frac{MG}{r^2}\left[1-\left(\frac{V}{c}\right)^2\right].
\end{eqnarray}
As it should be, this implies ${\bf a}\rightarrow 0$ as $V\rightarrow c$. In the Newton limit we have an expected
result, ${\bf a}=-\frac{MG}{r^2}$. At last, Eq. (\ref{rep14}) shows that the Schwarzschild field produces an attractive
force for any velocity $V<c$.

In conclusion, we specify these expressions for the case of a particle that reach infinity with zero velocity. Eq.
(\ref{rep11}) with $V_{\infty}=0$ gives the Drumaux's result for velocity, $V(r)=c\sqrt{\alpha/r}$. Concerning the
acceleration, resolving (\ref{rep11}) with respect to $V(r)$ and using the resulting expression in (\ref{rep14}) we
obtain
\begin{eqnarray}\label{rep15}
{\bf a}=-\frac{MG}{r^2}\left(1-\frac{\alpha}{r}\right)\left[1-\left(\frac{V_\infty}{c}\right)^2\right]
\quad\stackrel{V_\infty=0}{\longrightarrow}  \quad {\bf a}=-\frac{MG}{r^2}\left(1-\frac{\alpha}{r}\right).
\end{eqnarray}
Then magnitude of the acceleration is
\begin{eqnarray}\label{rep16}
a=\sqrt{g_{rr}}|{\bf a}|=\frac{MG}{r^2}\sqrt{1-\frac{\alpha}{r}}\left[1-\left(\frac{V_\infty}{c}\right)^2\right]
\quad\stackrel{V_\infty=0}{\longrightarrow}  \quad a=\frac{MG}{r^2}\sqrt{1-\frac{\alpha}{r}}.
\end{eqnarray}
The last expression should be compared with Drumaux's equation  (\ref{rep0}).

The sign of acceleration (\ref{rep15}) changes to the opposite when $r<\alpha$. But our basic definitions (\ref{rep9})
for $dl$ and $dT$ are only valid in the region $r>\alpha$, so the expressions (\ref{rep14})-(\ref{rep16}) can not be
applied inside the horizon.

%\section*{Acknowledgments}
{\bf Acknowledgments.} The work of AAD has been supported by the Brazilian foundation CNPq (Conselho Nacional de
Desenvolvimento Cient\'ifico e Tecnol\'ogico - Brasil),  and by Tomsk State University Competitiveness Improvement
Program.
%by the grant from The Tomsk State University D. I. Mendeleev Foundation Programm.
WGR thanks CAPES for the financial support (Program PNPD/2011).

%\begin{thebibliography}{00}

\end{document}